\documentclass[11pt,twoside]{article}

\def\chandra{{\it Chandra}}

\def\hst{{\it HST}}

\def\vla{{\it VLA}}

\def\lum{erg s$^{-1}$}
\def\flux{erg cm$^{-2}$ s$^{-1}$}   
\def\nh{cm$^{-2}$}
\def\arcsec{$^{\prime\prime}$}

\def\ltsima{$\; \buildrel < \over \sim \;$}
\def\simlt{\lower.5ex\hbox{\ltsima}} 
\def\gtsima{$\; \buildrel > \over \sim \;$}
\def\simgt{\lower.5ex\hbox{\gtsima}} 

\usepackage{asp2006} 
\usepackage{epsf} 
\usepackage{psfig} 
\usepackage{lscape} 
 
\begin{document} 
\title{An X-ray jet in the BL Lac S5 2007+777}   
\author{R. M. Sambruna, D. Donato, C.C. Cheung, F. Tavecchio, L. Maraschi}   
\affil{NASA/GSFC, NASA/GSFC, NRAO and Stanford, Oss. Merate, Oss. Brera}   
 
\begin{abstract} 
The BL Lac S5 2007+777 was observed by us with \chandra, to find the
X-ray counterpart to its 18\arcsec\ radio jet, and study its
structure. Indeed, a bright X-ray jet was discovered in the 33~ks
ACIS-S image of the source. We present its properties and briefly
discuss the implications. 

\end{abstract} 
 

\section{Introduction} 

One of the most important legacies of \chandra\ is the discovery and
the study of kpc-scale X-ray jets in Active Galactic Nuclei
(AGN). Mostly, \chandra\ detected X-ray emission from the jets of FRIs
and FRIIs \citep[see][for a review]{harr05}, raising important
questions of the mechanisms of emission of the high-energy radiation
and the acceleration processes in these sources.

The study of extended structure in BL Lac objects, however, lags
behind. X-ray emission from extended structures were so far detected
in only 3 of such sources: PKS~0521--365 (Birkinshaw, Worrall, \& 
Hardcastle 2001), 3C~371 \citep{joe01}, and PKS~2201+044 \citep{rms07a}. 
These objects are
peculiar, in the sense that broad optical emission lines were
occasionally detected from their nuclei, leading to an intermediate
classification between BL Lacs and QSOs. Indeed, their X-ray jets
present properties both of FRIs and FRIIs.

Here we report the discovery of an X-ray counterpart to the 18\arcsec\
radio jet in the BL Lac S5~2007+777 (z=0.342). The source is a
well-known BL Lac from the 1 Jy sample \citep{stick95} with a
narrow radio jet (Fig.~1). A more detailed discussion of the results
is in preparation \citep{rms07b}. 

\section{Chandra Observations and Results} 

We observed S5~2007+777 with \chandra\ on May 23, 2005. The ACIS-S
data were reducted according to standard procedures yielding 33~ks of
good data. Archival \hst\ (ACS F814W and WFPC2 F702W) and \vla\ 1.5
and 4.8~GHz observations were also used.

Figure~1 shows a montage of the multiwavelength images of the jet. At
radio, the jet appears knotty with several well-defined knots of
emission. 
X-ray emission is present from all the radio knots. Figure~1, bottom,
shows that there is a remarkable correspondence between the radio and
the X-ray emission. Table~1 lists the properties of the jet, including
the radio spectral index and the radio-to-X-ray broad-band index. The
X-ray flux and flux densities were calculated with \verb+PIMMS+ using
the knot's count rate in column 2 and assuming a power-law spectrum
with photon index $\Gamma=1.02$ and Galactic
N$_H$=8.58$\times10^{20}$\nh\ (Table~2). There are no variations of
the X-ray-to-radio flux ratio along the jet.

We extracted the X-ray spectrum of the full jet and of the brightest
knot at 8.5\arcsec\ from the core (Table~2). As for the latter a total
of 43 counts were collected, the spectral fit was performed using the
C-statistics. The full jet X-ray spectrum is well described by a power
law with $\Gamma \sim 1$. A thermal model is equally acceptable, but
the fitted temperature, kT=64 keV, effectively mimicks a power law. 
Similarly, knot K8.5 is fitted with a power law with rather hard slope 
(Table~2), albeit within large uncertainties. 

Table~2 also reports the spectral parameters from a fit to the core
X-ray spectrum. The latter is described by a power law with $\Gamma
\sim 2$, softer than the jet. Thus, as in the cases of 3C~371 and
PKS~2201+044, there is evidence for hardening of the X-ray emission going
from the core to the jet. No diffuse X-ray emission is detected around
the core. 

\section{Discussion} 

This paper presents new \chandra\ and archival \hst\ and \vla\
observations of the kpc-scale jet in the BL Lac S5~2007+777. An X-ray
counterpart to the radio jet was detected, with a remarkable
morphological correspondence at the two wavelengths. 

This morphological similarity provides first clues to the origin of
the high-energy emission. A likely possibility is that the X-rays are
produced via inverse Compton scattering of seed photons (CMB?) off the
same electrons responsible for the synchrotron radio emission. Indeed,
a similar 1:1 correspondence between the radio and X-ray jet is also
present in powerful FRIIs and similar or higher z, where IC/CMB is
thought to play a dominant role \citep{rms04}. Note that if
the X-rays were the high-energy tail of the longer-wavelength
synchrotron emission the size of the X-ray knots would be smaller than
at radio, due to the shorter lifetimes of the higher-energy
electrons, unless acceleration occurs throughout the entire volume. 

To better quantify the properties of the X-ray jet, we assembled its
Spectral Energy Distribution (SED) from radio to X-rays. Knot K8.5 was
used, as it is the brightest. A 3$\sigma$ upper limit to the optical
emission was derived from the ACS data. The SED is shown in Figure~2.
We also report the result of the modeling of the SED with the
synchrotron+IC/CMB model \citep*{fab00}. The hard X-ray emission can
be reproduced assuming that it belongs to the low energy tail of the
IC/CMB component: this choice implies a relatively large value for the
minimum energy of the relativistic electrons, $\gamma _{\rm min}=70$,
compared to the ``typical'' value $\gamma _{\rm min}=10-20$
\citep{rms02}. The bulk Lorentz factor is $\Gamma_L=15$. 

One of the motivations for observing this source with \chandra\ was to
test the structure of the jet. It has been proposed, and supported by
radio observations, that AGN jets may have two components - a fast
spine and a slow wall. The latter contributes synchrotron X-ray
emission when the jet is seen at larger viewing angles (as in FRIs),
while the fast-moving plasma in the spine is responsible for the
beamed IC/CMB emission seen in FRII jets.

BL Lacs are more aligned versions of FRIs. Thus, a significant
fraction of their jet X-ray emission should originate via IC/CMB off
the fast spine. S5~2007+777 was chosen for this study because of it
exhibits a long, collimated jet, while in most BL Lacs the radio jets
appear larger and more diffuse (Murphy, Browne, \& Perley 1993),
perhaps because of smaller viewing angles.

Our results confirm the original expectations that, at a favorable
angle, the X-ray emission from the jet of a BL Lac is dominated by
IC/CMB as in powerful FRIIs. This also provides circumstantial
evidence in support of a spine-wall structure for the jet.

\begin{table}
\caption{Knot flux/spectral properties}
\begin{center}
{\small
\begin{tabular}{l c c c c c c c} 
\tableline
\noalign{\smallskip}
Knot     &  Net c/s & F$_{\rm 0.3-8 keV}$ & F$_{\rm 1~keV}$ & F$_{\rm 4.86~GHz}$ & F$_{\rm 1.49~GHz}$ & $\alpha_{R}$ & $\alpha_{RX}$ \\
~~~(1)   &    (2)   &        (3)          &      (4)        &      (5)           &   (6)              &   (7)        &   (8)        \\

\noalign{\smallskip}
\tableline
\noalign{\smallskip}

K3.6     &   0.18   &   2.24              &   0.123         &   0.780            &   2.092            &      0.834   &     0.885    \\
K5.2     &   0.36   &   4.49              &   0.247         &   0.530            &   1.961            &      1.107   &     0.824    \\
K8.5     &   1.32   &  16.45              &   0.904         &   1.956            &   4.550            &      0.714   &     0.824    \\
K11.1    &   0.55   &   6.85              &   0.377         &   1.030            &   2.432            &      0.727   &     0.838    \\
K14.1    &          &		          &                 &                    &                    &              &              \\
K15.7    &   0.64   &   7.97              &   0.438         &   1.530            &   4.655            &      0.941   &     0.851    \\
K17.5    &          &		          &                 &                    &                    &              &              \\

\noalign{\smallskip}
\tableline                    
\end{tabular}
}
\end{center}
\scriptsize
{\bf Columns explanation}: 1=Knot name labeled as a 
function of the distance from the core (in arcsec) in the 1.49~GHz image; 
2=Net count rate in the 0.3--8 keV band in units of $10^{-3}$ c/s; 3=X-ray 
flux in the 0.3--8 keV band in units of $10^{-15}$ \flux;  4=X-ray flux 
density at 1 keV (in nJy); 5-6=Radio flux densities at the indicated 
frequencies (in mJy); 7=Spectral index in the radio band (between 1.49 and 
4.86~GHz); 8=Broadband spectral index between 1 keV and 4.86~GHz.

NOTE: The values for knot at 15.7\arcsec\ are obtained with an extraction 
region considerably large to increase the significance of the X-ray detection.
This area collects photons from all the 3 contiguous knots (i.e., K14.1
K15.7, and K17.5)


%
\end{table}

\normalsize
\vspace{0.2in}

\begin{table}
\caption{Results of the spectral analysis}
\begin{center}
\begin{tabular}{l c c c c c} 
\hline
Source   &  Net c/s &   $\Gamma$              & $\chi^2_r$/d.o.f. & F$_{\rm 0.3-8 keV}$ & L$_{\rm 0.3-8 keV}$  \\
   (1)   &    (2)   &      (3)                &      (4)          &      (5)            &   (6)                \\
\hline  													                    
CORE     &  23.59   &  1.98$\pm$0.23          &  0.89/201         &  246.65             &  90.555              \\ 
FULL JET &   0.28   &  1.02$^{+0.45}_{-0.43}$ &  1.38/3           &    4.30             &   1.308              \\
K8.5     &   0.12   &  0.84$^{+0.67}_{-0.69}$ &  ....             &    2.11             &   0.608              \\

\hline
                      
\end{tabular}
\end{center}
\scriptsize
{\bf Columns explanation}: 1=Source; 
2=Net count rate in the 0.3--8 keV band in units of $10^{-2}$ c/s; 
3=Photon index; 4=Reduced $\chi^2$ and degrees of freedom;
5=Observed X-ray flux in the 0.3--8 keV band in units of $10^{-14}$ \flux;  
6=Intrinsic X-ray luminosity in the 0.3--8 keV band in units of $10^{43}$ \lum.

\end{table}
\normalsize

\begin{figure}
\centerline{\psfig{figure=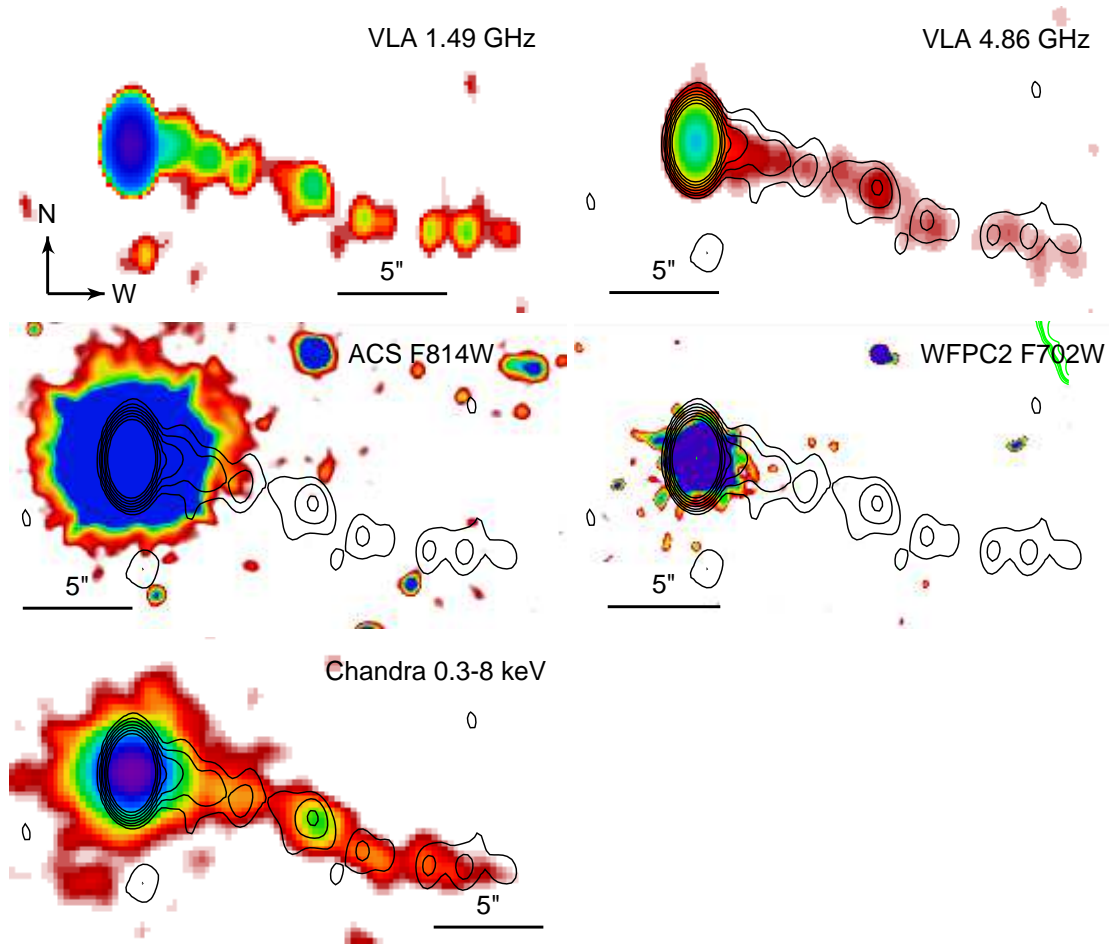,height=5.in}}
\caption{Images of the jet of S5~2007+777 at the various  
wavelengths. First row: \vla\ (1.49 GHz) and (4.86 GHz);  
Second row: ACS (F814W) and WFPC2 (F702W); Third row: \chandra\ 0.3--8 keV. 
In all cases, the 1.49 GHz radio contours are  
overlaid on the color image.}
\end{figure}



\begin{figure}
\centerline{\psfig{figure=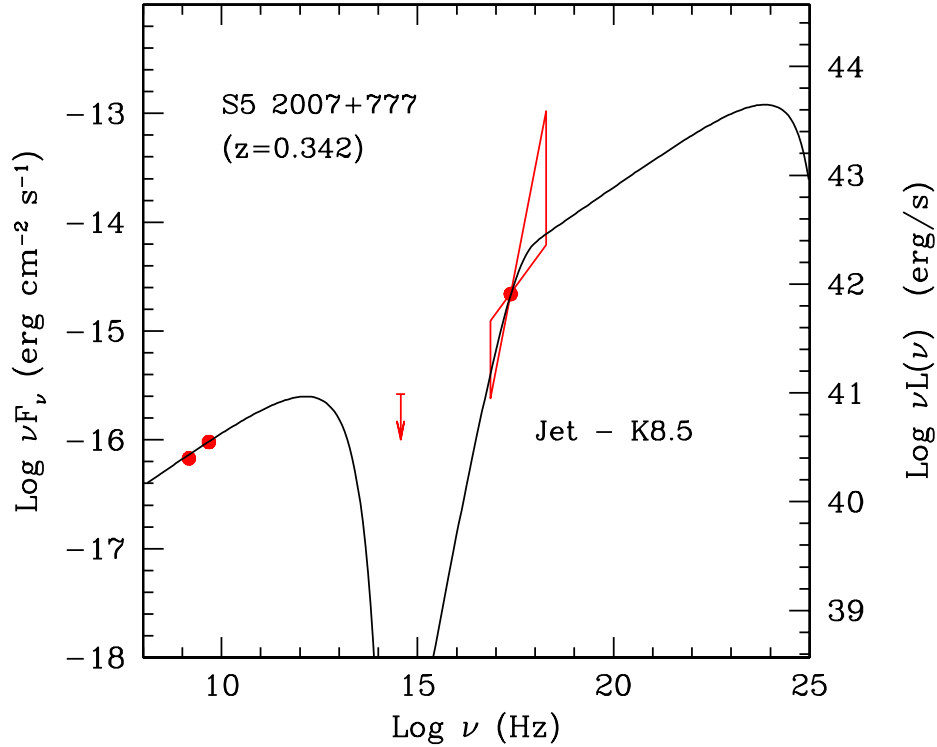,height=4.in}}
\caption{Spectral Energy Distribution of the brightest knot in the jet
of S5~2077+777. The solid line represents the IC/CMB model (see
text). }
\end{figure}

\end{document}